\documentclass{article}

\usepackage{arxiv}

\usepackage[utf8]{inputenc} 
\usepackage[T1]{fontenc}    
\usepackage{hyperref}       
\usepackage{url}            
\usepackage{booktabs}       
\usepackage{amsfonts}       
\usepackage{nicefrac}       
\usepackage{microtype}      
\usepackage{amsmath}
\usepackage{graphicx}
\usepackage{outlines}
\usepackage{subcaption}

\title{Augmenting Neural Differential Equations to Model Unknown Dynamical Systems with Incomplete State Information}

\author{
  Robert R. ~Strauss \\
  Los Alamos High School \\
  Los Alamos, NM 87544 \\
  \texttt{robert.strauss@studentlaschools.net} \\
}

\begin{document}
\maketitle
\begin{abstract}
Neural Ordinary Differential Equations \cite{FUNAHASHI1993801, chen2018neural} replace the right-hand side $F(.)$ of a conventional ODE with a neural net, which by virtue of the universal approximation theorem \cite{cybenko1989approximation}, can be trained to the represent any function. \cite{rackauckas2020universal,chen2018neural} It's inputs are the current state vector $\hat{x}$ and outputs are the derivative.
\begin{equation}
    \begin{aligned}
        \frac{\partial \hat{x}(t) }{\partial t} &= F[\hat{x}(t)]  \text{  and initial condition: } \hat{x}(0) = \hat{x}_0\\
    \end{aligned}
    \label{eq:0}
\end{equation} \\
If we knew the true, analytic, function $F(.)$ then we could directly train a neural net approximation directly to the output of $F(.)$ for varied inputs. But when we do not know the function itself, but just have the state trajectories (time evolution of the ODE system), we can still train the neural ODE to learn a representation of the underlying but unknown ODE. 

However, if the state vector at a given time, $\hat{x}$, is incompletely known then the right-hand side of the ODE is generally undefined. And thus the derivatives to propagate the system are unavailable, preventing training against the state trajectory data.

We show here that a specially augmented Neural ODE (ANODE) \cite{NIPS2019_8577} can learn the system model even when given incomplete state vector $\hat{x}$ information. As a worked example, we apply neural ODEs to the Lotka-Volterra problem of three interacting species: rabbits, wolves, and bears. We show that even when all the data for the bear population time series is removed, the remaining time series of the rabbits and wolves is sufficient to learn the dynamical system despite the missing incomplete state information and no knowledge of the ODE functional form. This is surprising since a conventional right-hand side ODE function cannot output the correct derivatives without the full state as the input.

A core, desirable, property of every ODE system, preserved by the neural ODE, is that trajectories in the state space cannot cross. The primary purpose of the ANODE architecture is not to predict the missing information but rather to enforce that even with missing state information. 

Surprisingly, we show that only by luck will the missing bear population be predicted correctly, even though the trajectory of the wolf and rabbit populations is being extrapolated correctly. Basically, the ANODE is finding some different neural ODE solution that has identical behavior to the true one on the known channels, but the functional form is not unique. 

Finally, we implemented the augmented neural ODEs and differential equation solvers in the \textit{Julia} programming language \cite{Julia-2017} within an existing neural ODE framework\cite{rackauckas2019diffeqfluxjl}. 

\end{abstract}
\section{Introduction}
\subsection{Problem}
The evolution of a system can be computed when encoded as an ordinary differential equation. Unfortunately, it is common to not know the functional form of a system of interest. By appeal to the universal approximation theorem\cite{cybenko1989approximation}, a neural network can be trained to approximate the ODE governing the system. Since the neural ODE\cite{FUNAHASHI1993801,chen2018neural} can be trained from data, a neural ODE can learn the rules governing a physical system with an unknown ODE functional form. 

However, an ODE cannot be evolved when the state of the initial state of the system is incompletely known. For example, the rate of change of populations in the Lotka-Volterra model of a predator-prey system depends only on the predator and prey populations; but if the number of predators is not known then the rate of change of the prey cannot be computed and so the system cannot be evolved in time. 

A complication occurs in training a neural ODE when the state trajectory is missing for some of the state variables. For example, if we know only the prey population time series but not the predator population then the training loss function won't demand the predator population be predicted correctly.

Intuitively, there is some reason not to despair. If we knew the exact functional form of the right-hand side $F(.)$, but not its parameters, then, in favorable cases, we might still be able to infer a varying predator population just by observing the prey population with time. That is, a falling prey population requires there exist a predator, and it falls faster the more numerous predators. Thus perhaps we might infer the proportional fluctuation of the predators from just the prey population dynamics even without knowing the initial condition.

A worse situation occurs when the reason we lack the data on a variable is that we don't even know the variable is a vital part of the state equations. For example, if we were not aware bears existed, a two-species ODE for rabbit and wolf population mechanics would be missing dependencies on an evolving bear predator population. This doesn't simply make the problem harder; it may entirely prevent a solution by a two-variable ODE. The feasible trajectory space of a two-variable ODE (rabbits and wolves) does not include every trajectory available to those same two variables when embedded in a three-variable ODE (rabbits, wolves, and bears).

This paper examines that hard challenge:  whether the system can still be solved when \underline{both} the entire functional form of the right-hand side is \underline{unknown} and, simultaneously, all of the training data for one state variable's trajectory is unknown. Furthermore, we restrict the training to short time series but require that it extrapolate to long-duration time series in testing. In each case study, we find the neural ODE successfully learns from just short time-series data then successfully predicts much longer time spans \textit{even} for initial conditions absent in the training set. The usefulness of neural ODEs for modeling unknown dynamical systems is thus demonstrated successfully.

\subsection{Relation to Prior Work}
Our principal work on this was completed in October 2019 and was subsequently published and presented in multiple, peer-reviewed, venues \cite{strauss2019jshs,strauss2020modeling,strauss2020countysf, strauss2020regionalsf}. In those we described our two novel Neural ODE architectures as "recurrent \textit{external} channel" neural ODEs to distinguish them from the conventional recurrent neural network (internal hidden nodes). As it happens, one of our two architectures is similar to one presented at NeuroIPS 2019 by Dupont \textit{et al}, but neither of us was aware of the other's contemporaneous developments  and so until now have not cited each other.\cite{NIPS2019_8577,dupont2019augmented}  In the interest of not fragmenting public terminology we now adopt Dupont \textit{et al}'s catchy acronym "ANODE" in this summary publication. The "recurrent \textit{external} channels" described in our previous work are logically the same as the augmentation variables "$\hat{a}(t)$" in Dupont \textit{et al}. 

Dupont \textit{et al} addressed a different problem than the one we posed. Moreover, that work set the undefined initial condition of the augmentation variables to a constant (zero) which was sufficient for their posed problem. But in general, this doesn't entirely resolve the non-crossing rule of ODE phase space trajectories that the ANODE (a.k.a. recurrent external channel) architecture was intended to address in the first place. In particular, it did not provide feasible solutions in the problems we posed for unknown systems (case-2 or case-3 below). To fully resolve this in case-3, we added a second neural net to compute unique initial conditions for the recurrent external channels that were conditioned on the data itself. Furthermore, this required that we trained both of these neural nets simultaneously. (See Methods section.)

For our case-1 (validation) experiment, we deliberately chose an identical ODE functional form and neural architecture to that of Rackauckas \textit{et al.} to conveniently verify our implementation. Surprisingly, we observed a different, significantly improved, outcome arising from a modest change in the training protocol. The changes (discussed below) appear to make the solution process more robust for somewhat stiff oscillatory equations and that protocol advancement was critical in the more challenging case-2 and -3 presented below.

\subsection{Background}

\subsubsection{Lotka-Volterra Equations}
As our concrete example, we shall use predator-prey population evolution modeled by the Lotka-Volterra equations \eqref{eq:1}. We refer to the species in this model as rabbits, which are prey only, and wolves, which eat rabbits. The canonical 2 species Lotka-Volterra equations are: \\
\begin{equation}
    \begin{aligned}
        \frac{\partial r}{\partial t} &= \alpha r - \beta r w \\
        \frac{\partial w}{\partial t} &= \delta r w - \gamma w
    \end{aligned}
    \label{eq:1}
\end{equation} \\
Where $w$ is the wolf population, $r$ is the rabbit population, and $\alpha$, $\beta$, $\gamma$, and $\delta$ are constants. \\

Adding Bears as an apex predator, eating both rabbits and wolves, the three species system becomes: \\
\begin{equation}
    \centering
    \begin{aligned}
        \frac{\partial r}{\partial t} &= \alpha r - \beta r w - \epsilon r b \\
        \frac{\partial w}{\partial t} &= \delta r w - \gamma w - \zeta w b \\
        \frac{\partial b}{\partial t} &= \eta r b + \theta b w - \iota b
    \end{aligned}
    \label{eq:2}
\end{equation} \\
Where $b$ is the bear population and all Greek letters are constants. \\

To train the neural ODE we need training and testing data sets in the form of evolving populations over time. We arbitrarily choose some "true" parameters $(\alpha,\beta,\epsilon,\delta,\gamma,\zeta,\eta,\theta,\iota )$ and then integrate the ODEs to generate the "true" population evolution over time. We will sample this continuous trajectory coarsely over a limited time range to generate the training data supplied to the neural ODE (which we may pretend are measurements from the field). (For purposes of plotting the "true" time series or the predicted time series we show continuous or finely sample wave-forms.) \\

After generating the data sets, no part of this "true" equation will be used in the neural ODE. None of the parameters of the true equation will be supplied to the neural ODE, nor will the functional form be given. 
The goal here is not to invert the neural ODE to an analytic function or to estimate the parameters of the "true" function. It is only to make a neural net that correctly extrapolates the system dynamics. Thus once we create the data the reader could safely forget they ever saw the true ODE.
This aligns with the target application of inferring neural ODEs for systems with unknown physics, for which the equations would, of course, not be provided like this. 

\subsubsection{Neural ODE}
This subsection is a quick primer just for readers unfamiliar with neural ODEs. A system can be simulated by integrating its differential equation. This means adding the rate of change of the state to the current state to get the future state, then repeating to produce a time-series of the state of the system. \\

To avert a common misunderstanding we note that the neural net does not directly predict the state of the system at a point in time in the future. It only acts as a replacement of the functional form of the ordinary differential equation, meaning a neural ODE must also be integrated through time to reach the future point \cite{sinai2019understanding}. Alternatively, one can view it as the asymptotic limit of a residual neural net with an infinite number of layers \cite{honchar_neural_2019, beer1995dynamics}. \\

At first, this seems counter-intuitive. Direct prediction would take a single evaluation passed through the neural network, while integrating may take thousands of evaluations, and so seems much less efficient. However, integrating through time produces not only produces better results but more importantly  preserves the non-crossing trajectory rule of ODE systems that is not preserved directly by a neural net on its own. Since physics happens to very often take the form of a differential equation this embeds strong restrictions on the possible outcomes. Enforcing this deterministic, rate-of-change, structure on the neural ODE makes it easier to learn the correct physics. Additionally, this form can enforce additional physical conservation laws. \\

One may ask, how is the neural net weight training accomplished when thousands of iterative passages through the integrator and neural net are needed to compute the trajectory used in the loss function? To backpropagate to the neural weight gradients requires the integrator itself to support automatic differentiation. \\

\section{Approach}
We aim to show neural ODEs can learn to model a system governed by unknown equations and missing state variables. For pedagogical illustration, we proceed in three stages of increasing complexity. First, a neural ODE is trained to match 2-species dynamics, then 3-species dynamics with one species's population evolution withheld from the training to simulate an unobserved variable in the state, and finally, 3-species dynamics with all state information about one species deleted. The first case is a well-known test problem, so it's here to introduce our notation and validate our process. It can be solved with a simple neural ODE containing a neural network with 2 inputs (the populations of the 2 species), giving two outputs, their respective rate of change. Novel augmented neural ODE architectures are required to solve the second two problems. These incorporate an augmented recurrent channel (Fig 3), and also a second external neural net to provide an initial condition (Fig 4). 

\begin{outline}[enumerate]
\1 Can a neural net predict how quickly wolves will eat rabbits?
\2 First, to test if a neural ODE really can learn to model a system without knowing an equation (just from data), we train a neural ODE to model a two-species predator-prey population system without giving it any knowledge of the equations.
\1 What if there are uncounted bears in the forest?
\2 Next, to test how the neural ODE handles an unmeasured variable, we add a third species to the system but hide the data for it, except for the initial population, from the neural ODE. 
\1 What if we had no idea there were bears in the first place?
\2 Finally, to test how a neural ODE handles a hidden variable problem, we fully hide the third species from the neural ODE and force it to model the other two species. It doesn't have any hint there even is a third species.
\end{outline}

In each case, the training is on the species populations "measured" over short time spans with coarsely sampled time intervals, then tested on a finely sampled longer time spans and with different initial conditions. The loss function is the total squared difference of the "measurements" (the time-series of the true populations) and the prediction (the time-series from integrating the neural network). Training is done in mini-batches of different initial conditions. \\

\subsection{Experiment 1: Just Wolves and Rabbits}
 The neural ODE is trained on population time series data with nothing withheld. This control will validate the neural ODE is sufficiently deep to reproduce the ODE dynamics and the training protocol is sufficient on short-time spans. Testing will compare the extrapolation accuracy on long time-span test-sets. \\

\subsection{Experiment 2: Hidden Bears}
In this case, the data for the bear population is withheld while training the neural network, \textit{except} we supply the true bear population at t=0 as the initial condition. The training loss function contains just the wolf and rabbit populations, ignoring the bear population. If you like, you may imagine it as a partially known variable -- you know bears are in a forest affecting the system, but they are only measured once at the start because it is too dangerous to the graduate students to repeatedly count bears in the woods.\\

One can see that this architecture has the capacity to correctly represent the differential equation since it has three inputs and three outputs. That is, if the neural network just happened to begin emitting the correct bear rate of change given the correct number of bear inputs, that's self-consistent and thus it has to work just as well as the previous model. But one might doubt it can be successfully trained to achieve that. First, since there's no constraint on the "bear" channel in the loss function it does not have to emit the correct bear population to achieve low loss. It could, for example, emit any monotonic transformation of the bear population (e.g. the cube root of the tangent of the bear population), because there are infinitely many possible ODE functional forms that are self-consistent with the wolf and rabbit population trajectories. Furthermore, this ambiguity might make it too hard to successfully learn any model at all without the helpful constraint of the withheld bear population evolution in training data. \\

The ODE neural net has three inputs. Two of these are the wolf and rabbit population. Intuitively, we would like to think of the third one as "bears". But as we shall see it does not turn out to be the bear population. It's just an additional recurrent external channel. There are three outputs which provide the rate of change for the Rabbit and Wolf and recurrent external channel to the integrator.\\

\subsection{Experiment 3: What Bears?}
This is almost the same as the previous test, except the external channel is not initialized to the bear population at t=0. We devised another approach to initializing it. Simply supplying a constant, like zero, would not be likely to generalize to a correct result on testing data. This can be seen by imagining two test data sets generated starting with the same initial number of rabbits and wolves but different numbers of bears. The different number of bear predators means the trajectories must be different, but every trained ANODE will have a single deterministic output for the same three inputs (wolves and rabbits and zero). Thus it could not match both data sets. This is a special case of the general rule that ODEs cannot produce crossing trajectories. 

Thus more than just the ANODE architecture is needed to resolve this. We implemented a separate neural network outside of the neural ODE which receives not just the t=0 wolf and rabbit populations but the first ten time points of the wolf and rabbit population. It outputs a single scalar to initiate (t=0) the recurrent external channel. Thus the initial conditions delivered to the neural ODE itself are the ordinary t=0 values of the rabbit and wolf, as well as this derived initial condition for the recurrent channel. \\
Note that both of the initial-condition-predictor and the neural net inside the ODE must be trained simultaneously. We are not separately training the external net to predict some known bear population;  we would not have that initial bear information since we are pretending we don't know there are bears at all. This means we also back-propagate through both neural nets including through all iterations of the solver and neural ODE to further back-propagate through the initial condition determining network.

\begin{figure}[htbp]
    \caption{ \textbf{Training (Wolves and Rabbits)} \textit{All}: x-axis is time, y-axis is population. Dots represent the prediction from integrating the neural ODE, lines represent the true solution to the Lotka-Volterra equations. There are four plots, each made after different amounts of training. By the end the dots align well with the lines, meaning after training the neural ODE learns to accurately predict the true solution.}
    \begin{subfigure}{0.49\textwidth}
        \centering
        \includegraphics[width=0.9\textwidth]{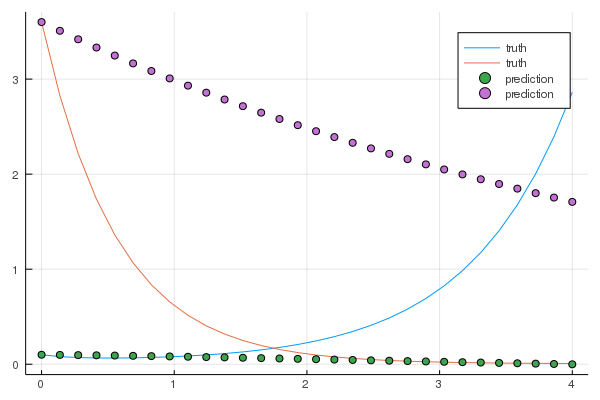}
        \caption{Prediction before training, compared to truth}
    \end{subfigure}
    \begin{subfigure}{0.49\textwidth}
        \centering
        \includegraphics[width=0.9\textwidth]{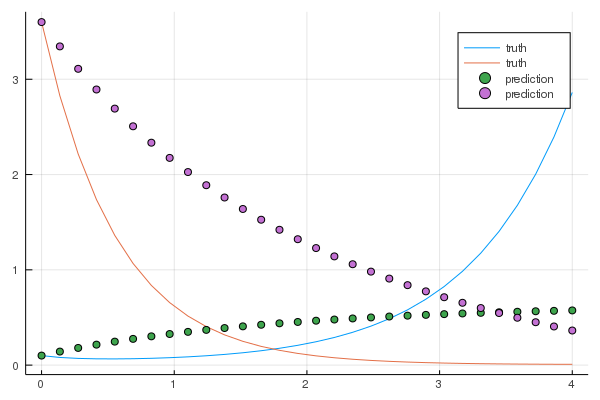}
        \caption{Prediction after some training, compared to truth}
    \end{subfigure}
    \begin{subfigure}{0.49\textwidth}
        \centering
        \includegraphics[width=0.9\textwidth]{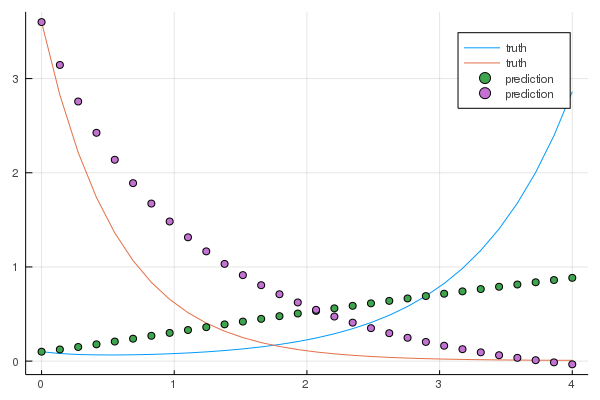}
        \caption{Prediction after more training, compared to truth}
    \end{subfigure}
    \begin{subfigure}{0.49\textwidth}
        \centering
        \includegraphics[width=0.9\textwidth]{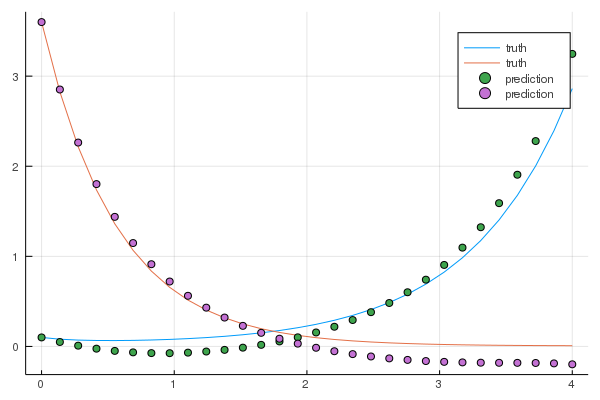}
        \caption{Prediction after even more training, compared to truth.}
    \end{subfigure}
    
    \label{fig:training}
\end{figure}

\section{Results}

\subsection{Observing the Training}
Through training, the prediction from integrating the neural ODE goes from uniformed and inaccurate to matching the true solution of the Lotka-Volterra equations. The gradual progression of training to match the Lotka-Volterra equations is shown in figure \ref{fig:training}. To test if the neural network actually learned to generalize rather than memorize, the model is applied to an initial condition it hasn't trained on and, even more stringently, tested over a longer time span, so it must extrapolate using the $F(.)$ it has learned. The plots in figure \ref{fig:data} are the results from these tests in each experiment.
\subsection{Experiment 1: Just Wolves and Rabbits}
To test what it learned, the neural ODE and Lotka-Volterra equations are solved with an initial condition not seen in the training set and over a longer period of time than in training. Figure \ref{fig:wr} shows the prediction from the neural ODE matches up with the true solution to the Lotka-Volterra equations in these testing conditions very well. This shows the neural net was successfully trained through back-propagation through the ODE solver. Given it extrapolates to the held-out testing data establishes the neural network learned the physics of the system rather than memorizing the test data. This demonstrates neural ODEs are able to learn the physics of unknown systems if no data is withheld.\\ 
\subsection{Experiment 2: Hidden Bears}
Again, to test what it learned, the neural ODE and Lotka-Volterra equations are solved with an initial condition not seen in the training set and over a longer period of time than in training. Figure \ref{fig:wrb} shows the prediction from integrating the neural ODE matches up with the true solution to the Lotka-Volterra equations in only two of the three species. This is confusing at first. How can the neural ODE model the rabbits and wolves without knowing the bear population?  \\

We were initially surprised that the bear population did not match the recurrent external state variable. It seemed like that since we gave it the true bear population as the initial condition that the simplest outcome would be to have reproduced the original ODE. But it did not. And it didn't have to because the loss function did not impose this. So it doesn't need to predict the bear population, the recurrent channel only needs to transmit and update some information that aids it in predicting the other two populations. However, it is hard to imagine a general class of transformations that also would be self-consistent with every initial value matching the bear population at t=0. Thus we expected it would be biased to find the true bear population, and we were surprised when it found some alternative solution. 

\subsection{Experiment 3: What Bears?}
\begin{figure}[htbp]
\caption{
    \textit{All}: X-axis is time, y-axis is population, lines represent the solution of the Lotka-Volterra equations, dots represent the prediction from integrating the neural ODE. \\
    \textit{\ref{fig:wrb}, \ref{fig:wrwb}}: Blue is rabbits, orange is wolves, green is bears, purple dots are predicted rabbits, yellow dots are predicted wolves, blue dots are predicted bears/recurrent variable. \\
    \textit{\ref{fig:wr}}: Blue is rabbits, orange is wolves, green dots are predicted rabbits, purple dots are predicted wolves.
}
\begin{subfigure}{0.49\textwidth}
    \begin{center}
        \includegraphics[width=\textwidth]{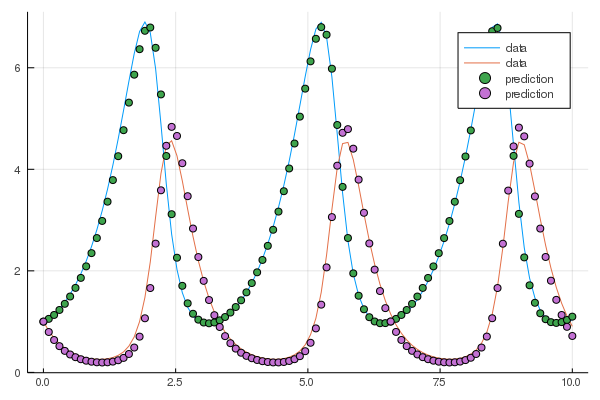}
    \end{center}
    \caption{\textbf{Testing (Wolves and Rabbits)} The prediction from integrating the neural ODE matches the true solution of the Lotka-Volterra equations.. This shows neural ODEs can learn the physics of a system and extrapolate to other initial condition, when provided enough data to train on.}
    \label{fig:wr}
\end{subfigure}
\hfill
\begin{subfigure}{0.49\textwidth}
    \begin{center}
        \includegraphics[width=\textwidth]{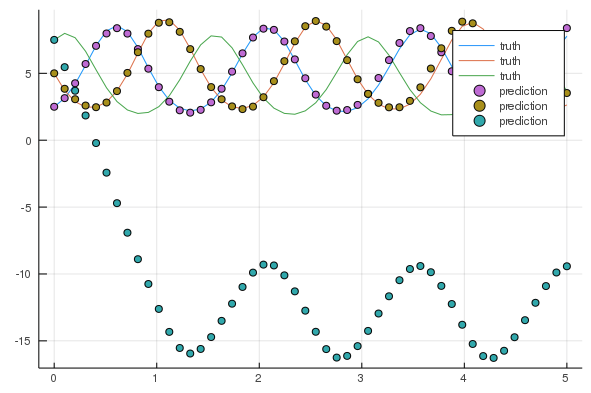}
    \end{center}
    \caption{\textbf{Testing (Hidden Bears)} Two of the three dotted lines match with the lines, while the third dotted line has a different behavior. This shows neural ODEs can learn to model some variables of a system, without training on data on other variables}
    \label{fig:wrb}
\end{subfigure}
\hfill
\begin{subfigure}{0.49\textwidth}
    \begin{center}
        \includegraphics[width=\textwidth]{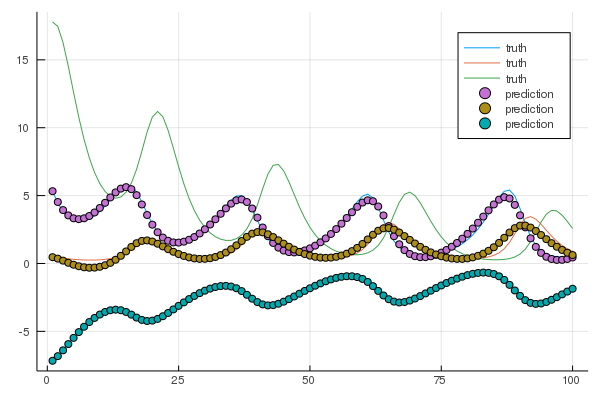}
    \end{center}
    \caption{\textbf{Testing (What Bears?)} Two of the three lines match, while the third dotted line is far from the corresponding line. This shows neural ODEs can learn to model some variables of a system while other components are completely hidden from it in training. }
    \label{fig:wrwb}
\end{subfigure}
\label{fig:data}
\end{figure}

 Figure \ref{fig:wrwb} shows the prediction from integrating the neural ODE matches up with the true solution to the Lotka-Volterra equations in two of the three species, and has different behavior in the third. Again, only the rabbit and wolf populations match, and they match very well.  Notice that, unlike the previous case, it now does not match the initial value of the bears. It does seem to have oscillatory dynamics like the true bear population but magnitudes and phases evolve differently over time.

\section{Discussion and Conclusions}

We have shown that, not only can a neural ODE be trained to model systems with unknown physics from just measurements, but it can resolve cases where the system has 1) known but immeasurable variables or 2) depends on unanticipated and unmeasured state variables. While this has the suspicious aura of magic, not physics, it is simply because knowing something \textit{can be} modeled by an ODE, even without knowing the ODE, imposes very strong constraints on the outcome. The neural ODE formulation is thus an efficient and highly adaptable means of learning the underlying model from data while automatically filtering out all ODE-incompatible models. This achieves something that otherwise might be incomprehensible or defy easy parameter estimation techniques. \\

Our case 1 was intentionally patterned after the Lotka-Volterra neural ODE studied by Rackauckas \textit{et al.} to be able to compare the results. Rackauckas \textit{et al.} investigated the problem of reverse engineering to a simple analytic functional form from a trained neural ODE. They trained on short-time data sets as well and found that the reverse-engineered analytic function could extrapolate to longer time spans far better than their trained neural ODE. Intriguingly, while their trained neural ODE extrapolated poorly, our identical architecture neural ODE extrapolated very well for all the parameter variations we tried. We speculated subtle training differences might matter more than one might expect, including 1. small changes in the number of time points in the interval, 2. small changes in the time span (ours was the typical interval between population spikes), and 3. the use of batch training from many different initial conditions. \\

Tentatively exploring these, we found only the last of these had a profound impact. Training by serial stochastic gradient descent usually terminated on quite poor local minima. In contrast, mini-batch training was excellent. Counter-intuitively, we observed that giving the training process data over longer time spans produced even worse results. Intuitively, spans covering many oscillatory spikes should have been more informative, just as Fourier estimation works better with time ranges longer than a period. But evidently, they are practically harder for gradient methods to optimize in the context of an ODE integration process. Instead when those long intervals were split up into short interval mini-batches the training was efficient and converged to better solutions. That is, short time spans of spiky time series are much easier to optimize but perform poorly in extrapolation, but mini-batches allow gradients on all different phases of the population cycle at once enabling both easier fitting and better extrapolation. Thus the most critical difference to the extrapolation was not simply the total amount of data we used compared to Rackauckas \textit{et al.} (who used less) but the division of long multi-cycle time series into batches of short time series. We think this is an unexpected and valuable practical finding. (Chris Rackaucus has now updated the neural ODE optimization API and added documentation to the \texttt{DiffEqFlux.jl} library to facilitate batch learning. Unfortunately, our code, available on GitHub, predates that so it doesn't demonstrate this nice new API.) 

In cases 2 and 3, the values of the recurrent external state variable did not match the bear truth data. The functional form the neural net has learned is \underline{not} the canonical Lotka-Volterra equations. For example, in the canonical form there a simple quadratic monomial (wolves times bears) to represent the loss bears impose on wolves. This could be replaced by other ways of coupling these state variables when we have the freedom to make the external state variable be anything we want. For example, one could make this channel be tangent(bears*wolves+rabbits) and there still would exist some ODE set that would correctly propagate the populations of wolves and rabbits. So the neural net isn't locked into using the same state variables nor estimating the same functional form used to generate the data. \\ 

Thus, except by chance, the third channel should never give the original "bear" population. Even when we biased the system towards being bear-like in case 2 (by supplying the "bear" initial condition) it still did not cause the external channel to discover a state variable that followed the bear population. It's simply an abstract external channel that provides recurrent information necessary to make this a 3-variable, not a 2-variable state space.\\

This raises an intriguing set of questions that are beyond the scope of this work. Namely, does the additional freedom unleashed by the infinite number of possible solutions it might learn to make the training process "easier" or "harder"?  Given stochastic descent as the training, will random mini-batch orders lead to different solutions or are there attractors that recur frequently?  If one reverse engineered many different functional forms of the implicit ODE in the manner of Rackauckas \textit{et al.} would they all belong to some easily described transformation group?\\

\section{Methods}

\subsection{Network Design}
Our new neural network architecture was designed specifically to address the central problem arising from incomplete state specifications. It preserves the core property of all stationary ODEs: trajectories don't cross. The crossing of trajectories does not occur in the fully specified ODE of known systems simply because the next integration step along the time derivative $\frac{\partial \hat{x}(t) }{\partial t}$ depends only on the current position $F[\hat{x}(t)]$. Two ramifications occur for unknown systems. First, if one is unaware of an additional member variable in an N-dimensional $\hat{x}$, then the reduced state space of N-1 dimensions can have a position that maps to many states in the N-dimensional space with different gradients, and so paths cross. Second, even if one knows the problem is N-dimensional if the initial condition isn't specified for all the elements of $\hat{x}$, the same problem arises. \\

\begin{figure}[ht]
\caption{Diagram of the structure of first new augmented neural ODE. Wolf and rabbit populations are input to the neural ODE and their respective rate of changes are output. The third variable (previously symbolizing bear population) is also input, and its rate of change also output.}
\centering
\includegraphics[scale=0.4]{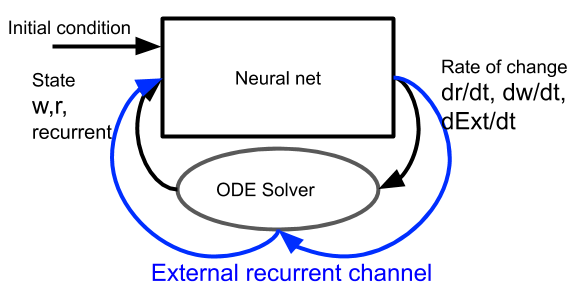}
\end{figure}

The first problem is solved by augmenting an N-1 dimensional ODE with an additional idler variable meant to make this mapping unique again. We may not know what this variable is, but we just add it on. This need not be done blindly. We need only need to add as many variables as required to remove trajectory crossings in the training data. \\

\begin{figure}[h]
\caption{Diagram of the structure of the second new augmented neural ODE. Neural net 1 takes in the first 10 data-points in wolf and rabbit populations and outputs initial condition for the third variable. (This is not trained to output the exact bear population, it is trained in conjunction with the neural ODE so only guesses a useful number, not the real bear population.) }
\centering
\includegraphics[scale=0.6]{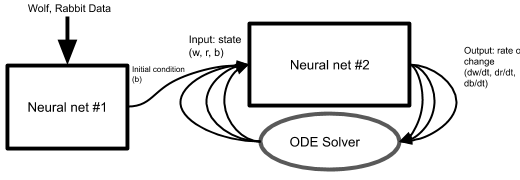}
\end{figure}

Secondly, how do we initialize these new variables or any other variables for which we lack that initial data?  We address this by training a second neural net which takes in a time series of the variables whose trajectory we do know, and outputs a unique initial condition. Thus the trajectory ambiguity of the unknown initial condition is removed and again paths don't cross. We note that this degeneracy could be broken by any arbitrary mapping of known data to a unique initial condition (e.g. a hash function) but such would be unlikely to have nice continuity properties and thus requires a deeper neural net to handle the arbitrary discontinuity. Therefore, we used a neural net to more smoothly estimate the initial condition. \\

\subsection{Implementation}
We implemented the augmented neural ODEs and differential equation solvers in the \textit{Julia} programming language \cite{Julia-2017} within an existing neural ODE framework\cite{rackauckas2019diffeqfluxjl}. This was all completed by writing code in the \textit{Julia} programming language version 1.0. The truth ODE function for the Lotka-Volterra equations was written and solved in each case for truth data. Neural networks were created with the \texttt{Flux} package \cite{Flux.jl-2018}. Differential equations were solved with the \texttt{DifferentialEquations} package \cite{DifferentialEquations.jl-2017}, and back-propagation through a neural ODE used the \texttt{DiffEqFlux} package \cite{rackauckas2020universal}.\\
The neural ODE network structure was intentionally kept the same as ones used for solving the canonical Lotka-Volterra problem with 3 dense hidden layers, 32 nodes wide, using swish activations. These had 2 (case 1) or 3 (cases  2\&3) inputs/outputs. The structure of our initial condition neural net was 20 inputs (10 each for r and w), one output (the initial condition for the external channel), with 3 hidden dense layers 32 nodes wide with swish activation. Training used the Flux.jl package \cite{Flux.jl-2018} with ADAM optimization. The code and documentation written for this project can be found online at \url{https://github.com/robertstrauss/nnDiffEq}.

\section{Frequently Asked questions}
For the benefit of readers less familiar with neural ODEs this subsection will raise and answer a few straw-man issues that we have been asked about many times at presentations of this work.\\  

\underline{The enabling innovations of this work} are the "recurrent ODE channel" and the process of initializing it. Crucially, this is distinct from a "recurrent channel" in the conventional neural network sense of the term "recurrent". Unlike what is normally meant by a recurrent node, this "recurrent channel" actually passes through the ODE solver too and is integrated.\\ Besides imposing the ODE trajectory rules on the channel there is also a crucial practical reason we cannot (easily) use a conventional, non-integrated, recurrent neural network internal state. The ODE solver algorithm itself might jump back and forth in time (e.g. adaptive step size, successively approximating, backward-forward algorithm). By letting the channel be differential we let the solver itself handle the intricate bookkeeping of the time spacing and order of its calculations, just as it does for integrating the other normal channels. For clarity, we note that for the special case of a simple, fixed-time-step, forward Euler integrator, one could in principle get away with the recurrent channel being a conventional hidden internal state instead of having it output the derivative of the channel to be passed through the integrator. But in that trivial case, there would also be no need to pass any of the channels through the integrator, since a recurrent neural net could also be tasked with learning the Euler integration itself.\\

Since the use of an ODE integrator seems like it adds complexity, one might ask \textit{Why not just have a neural net directly predict the entire solution instead?}  Surprisingly, the ODE actually simplifies the process while it also ensures the solution will have the properties we expect. An ODE forces the curves to be continuous and non-crossing no matter how sparsely sampled (not enforced by an ordinary neural net). Thus we are exploiting the prior knowledge that there is some way of describing the system with an ordinary differential equation even if we don't know that ODEs or even all the variables of the system. Similarly, this can enforce conservation laws. Although in this example there is no conservation law between wolf and rabbit populations, in other cases this becomes important. It's simpler because it's recurrent: one is using the same neural net weights for every output node of the trajectory no matter when they occur in the time series, so there will be far fewer weights to train. Being smaller it will be easier and faster to manage the optimization. And it can ensure the solution obeys consistent behaviors. For example, in a stationary ODE, if all the free variables in a system return to a set of values that they already had, the same behavior should follow now as when it previously happened. If something happens twice, expect the same outcome each time regardless of when it occurs. Direct prediction from a recurrent neural network can violate this principle. \\

\textit{Why didn't we use a parameter-fitting model containing dozens of guesses at possible mathematical combinations of the input (such as a Taylor series) rather than an entire neural network to find the rule?} Parameter fitting seemingly has many advantages: much fewer trainable parameters (a parameter fitting model might have a few dozen while a neural net has thousands), and a readable outcome telling you which terms were needed and thus giving the functional form at the end. Recent research has shown that directly curve fitting to an ensemble of possible functional forms just does not seem to be as robust.\cite{rackauckas2020universal}. It is unclear why this is, but for now, it appears the best strategy for model inference with limited or noisy data is to train the neural ODE first then use it to generate additional data for parameter fitting to an ensemble of functions.\cite{rackauckas2020universal} 

The third channel is absolutely required since it would be impossible for a neural ODE to predict correctly with only the input of two of the three populations and no other information because the correct output depends on all three populations. That is for the same rabbit and wolf population the derivative output is a multi-valued function that will differ depending on the bear population. We create the recurrent ODE channel to give the neural ODE more information, in the hope that will be enough to allow it to predict correctly. This setup makes logical sense because, for a system for which we lack equations, we may not even realize a variable at play in the system. This "recurrent channel" acts as an open slot for variables which we do not know about, but are required for the prediction of the system. \\

\section{Acknowledgements}
Thanks to Dr. Charlie Strauss for mentoring and editing. Thanks to Chris Rackauckas for his inspirational online video tutorials and early access to his pre-print on Universal Neural ODEs.\cite{rackauckas2020universal}


\end{document}